\documentclass[letters, usenatbib]{mnras}
\usepackage{amssymb,amsmath}
\usepackage{graphicx}
\usepackage{xcolor,natbib}
\usepackage{multirow}
\usepackage[T1]{fontenc}
\usepackage{ae,aecompl}
\usepackage{newtxtext, newtxmath}
\usepackage{float}
\usepackage{subfigure}
\usepackage{longtable}
\usepackage{enumitem}
\usepackage{longtable,listings}
\usepackage[flushleft]{threeparttable}
\usepackage{parcolumns}
\usepackage{makecell}

\def\oc3{[O~{\sc iii}]$_c$}
\def\ob3{[O~{\sc iii}]$_b$}
\def\obj{SDSS J1605}

\title[A TDE candidate in double-peaked BLAGN]
{A candidate for central tidal disruption event in the broad line AGN SDSS J1605 with double-peaked broad H$\beta$}
\author[Zhang]
{Xue-Guang Zhang\thanks{Corresponding author
    Email: \href{mailto:xgzhang@gxu.edu.cn}{xgzhang@gxu.edu.cn}}\\
    Guangxi Key Laboratory for Relativistic Astrophysics, School of Physical Science and Technology,
    GuangXi University, Nanning, 530004, P. R. China}

\pubyear{2023}

\begin{document}

\label{firstpage}

\pagerange{\pageref{firstpage}--\pageref{lastpage}}

\maketitle

\begin{abstract} 
	In this letter, motivated by double-peaked broad Balmer emission lines probably related to tidal disruption events (TDEs), 
a potential TDE candidate is reported in SDSS J160536+134838 (=SDSS J1605) at $z\sim0.44$ having quasar-like spectrum 
but with double-peaked broad H$\beta$. The long-term CSS light curve can be naturally described by a main-sequence star of $2.82_{-0.19}^{+0.20}{\rm M_\odot}$ disrupted by the central black hole (BH) of $144_{-21}^{+26}
\times10^6{\rm M_\odot}$ in SDSS J1605. Meanwhile, the ASAS-SN light curves afterwards show none apparent trend variability, 
indicating the bright CSS flare in SDSS J1605 unique and different enough from variability of normal AGN. Furthermore, there is a 
consistency between the TDE model determined sizes of debris with the sizes of emission regions for the double-peaked broad H$\beta$ 
described by the accretion disk model, supporting the disk-like BLRs probably related to a central TDE in SDSS J1605. And the virial 
BH mass $\sim$7 times higher than the TDE model determined value can be naturally explained by R-L relation determined BLRs sizes 
very larger than the actual distance of emission regions related to TDEs debris in SDSS J1605. Although no clear conclusion on 
double-peaked broad lines absolutely related to TDEs, the results here provide clues to detect potential TDEs in AGN with 
double-peaked broad lines.
\end{abstract}


\section{Introduction}

	Tidal Disruption Events (TDEs) have been studied for more than four decades \citep{re88, lu97, gm06, gr13, gm14, wy18, 
mg19, tc19, ry20, lo21, zl21}, with accreting fallback debris from stars tidally disrupted by central supermassive black holes 
(SMBHs) leading to apparent time-dependent variability. More recent review on theoretical TDEs can be found in \citet{st19}. 
TDEs are commonly accepted as the better indicators to SMBHs and BH accreting systems, more and more TDEs have been detected 
and reported in the literature (see \url{https://tde.space/}), especially through current public sky survey projects.

	Descriptions of one to two known TDEs candidates from each current survey are given as follows. Through the 
SDSS Stripe82 database, \citet{ve11} have reported two TDEs candidates in SDSS non-active galaxies. Through the Catalina Sky Survey 
(CSS), \citet{dd11} have reported a probable TDE candidate in narrow-line Seyfert 1 galaxy CSS100217. Through the project of Panoramic 
Survey Telescope And Rapid Response System (Pan-STARRS), \citet{gs12, ch14} have reported robust TDEs candidates of PS1-10jh and 
PS1-11af in inactive galaxies, due to their ultraviolet-optical flares well described by the theoretical TDE model. Through the 
project of Palomar Transient Factory (PTF), \citet{bl17, vv19} have reported two TDEs candidates of iPTF16fnl and AT2018zr/PS18kh. 
Through the project of Optical Gravitational Lensing Experiment (OGLE), \citet{wz17, gr19} have reported TDE candidates of OGLE16aaa 
and OGLE17aaj. Through the project of All-Sky Automated Survey for SuperNovae (ASAS-SN), \citet{ht14, ht16, hi21} have reported the 
known TDEs candidates of ASASSN-14ae, ASASSN-14li and ASASSN-19dj. Through the Caltech-NRAO Stripe82 Survey (CNSS), \citet{an20} 
have reported the TDE candidate of CNSS J0019+00. Through the project of Zwicky Transient Facility (ZTF), \citet{lh20, sv21} have 
reported the TDE candidate of AT2019DSG. Through the project of the Global astrometric interferometer for astrophysics 
(Gaia), \citet{kr18} have listed a small sample of TDEs candidates among the nuclear transients reported by the Gaia Science Alerts 
team. Through the project of the Asteroid Terrestrial Impact Last Alert System (ATLAS), \citet{ht23} have discovered the TDE candidate 
ATLAS18mlw, a faint and fast TDE in a quiescent galaxy. More recently, two large samples of dozens of new TDE candidates can be 
found in \citet{vg21} from the First Half of ZTF Survey observations along with Swift UV and X-ray follow-up observations and in 
\citet{sg21} from the SRG all-sky survey observations and then confirmed by optical follow-up observations. More recent review on 
observational properties of the reported TDEs can be found in \citet{gs21}.

\begin{figure*}
\centering\includegraphics[width = 18cm,height=8cm]{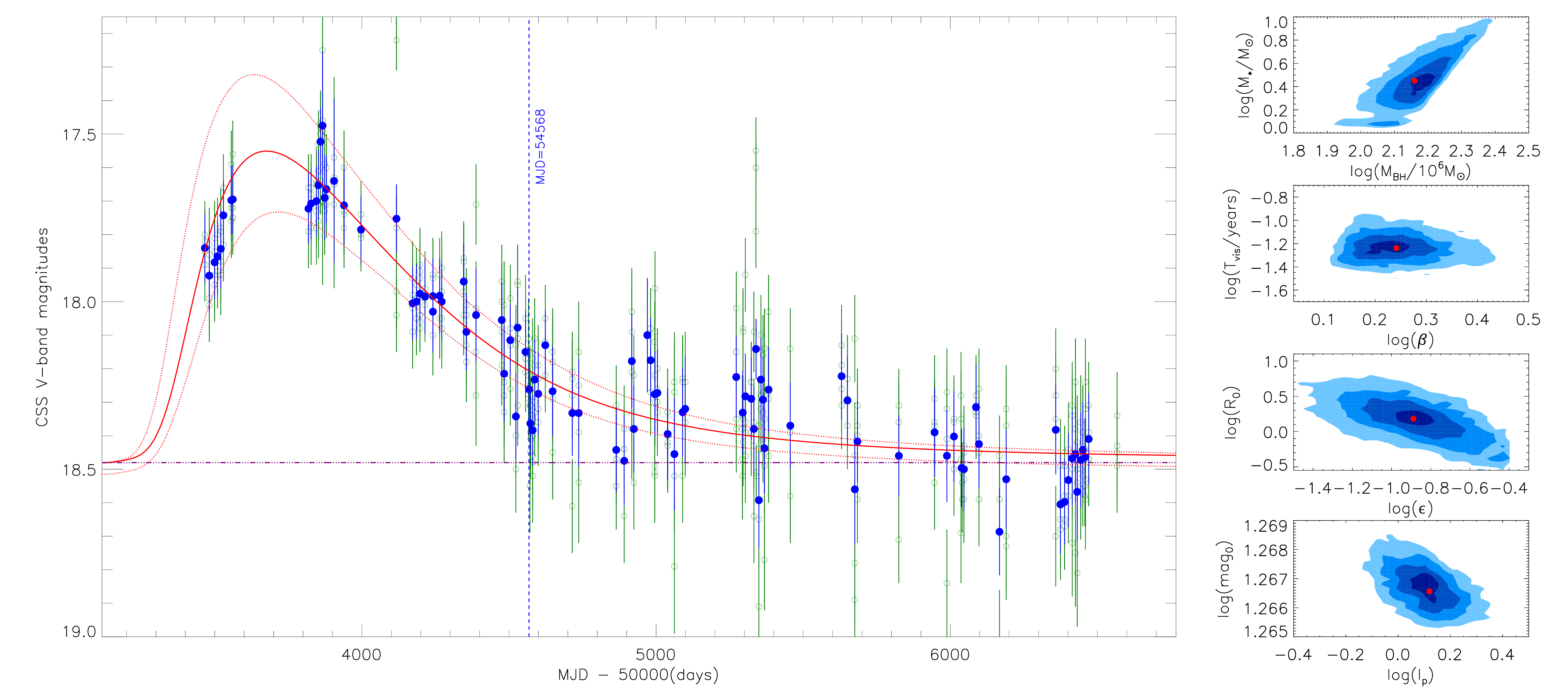}
\caption{Left panel shows the best descriptions (solid red line) to the CSS V-band light curve (open circles plus error bars in 
dark green) by the theoretical TDE model expected variability plus a none-variability component. Solid circles plus 
error bars in blue show the corresponding 5days binned light curve. Dashed red lines show the confidence bands to the best 
descriptions, after accepted the uncertainties of the TDE model parameters. Horizontal purple line shows the determined none-TDE 
component with magnitude about 18.48. The vertical dashed blue line marks the position MJD=54568 on which the SDSS spectrum is 
observed. Right panels show the MCMC determined two-dimensional posterior distributions in contour of the TDE model parameters and 
$mag_0$. In each right panel, the five contour levels show the corresponding 0.1, 0.3, 0.5, 0.7 and 0.9 of the 
two-dimensional volume contained, and solid red circle marks the positions of the accepted values of the model parameters.}
\label{lmc}
\end{figure*}

	Among the reported TDEs candidates, it can be commonly accepted that weak narrow emission lines (especially weak 
[O~{\sc iii}]$\lambda5007$\AA~ emission lines) but broad Balmer and He~{\sc ii} emission lines are in their optical spectra. In part of TDEs candidates, the reported broad emission lines could be related to disk-like structures from TDE 
debris, such as SDSS J0159 \citep{md15, zh21}, PTF09djl \citep{lz17}, PS18kh \citep{ht19}, AT2018hyz \citep{sn20, hf20}, 
AT2019qiz \citep{sl23}, etc., indicating the reported optical broad emission lines in part of TDEs candidates  
are not related to normal broad emission line regions (BLRs) in broad line AGN (BLAGN) but tightly related to TDE debris. In 
other words, potential TDEs candidates could be probably detected in BLAGN with double-peaked broad Balmer emission lines 
(double-peaked BLAGN). Motivated by the point that double-peaked BLAGN could possibly harbor potential TDEs, the SDSS 
J160536+134838 (=SDSS J1605) at redshift $z\sim0.44$ is collected as the subject of this letter, not only due to 
its broad Balmer lines described by a disk-like component but also due to its TDE expected photometric variability. Not similar 
as the previous reported TDEs in nearby inactive galaxies and/or in weak active galaxies and/or in changing-look AGN, the SDSS 
J1605 harboring a potential central TDE reported in this letter has not only strong and apparent broad Balmer emission lines 
but also strong [O~{\sc iii}]$\lambda5007$\AA~ emission lines.

	This letter is organized as follows. Section 2 presents main results on photometric variability properties, and necessary 
discussions. Section 3 shows main results on broad emission lines and necessary discussions. Section 4 gives final conclusions. 
And in the Letter, we have adopted the cosmological parameters of $H_{0}=70{\rm km\cdot s}^{-1}{\rm Mpc}^{-1}$, 
$\Omega_{\Lambda}=0.7$ and $\Omega_{\rm m}=0.3$.

\section{Long-term Variability in SDSS J1605} 

	Long-term photometric V-band light curve of SDSS J1605 with $z\sim0.44$ is collected from CSS \citep{dr09} with MJD from 
53466 to 56567 shown in the left panel of Fig.~\ref{lmc}. Meanwhile, the corresponding high-quality 5days binned light 
curve is also shown. There are two points we can find. First, in the 4 years with MJD from $\sim$5200 to $\sim$6600, there are not 
apparent trend variability in SDSS J1605. Second, there is a smooth declining trend with MJD from $\sim$3800 to $\sim$5000, 
similar as the variability pattern expected from a TDE. Here, the trend variability amplitudes are about 0.75mag and 
0.02mag simply determined by a linear function applied to the light curve (also the binned light curve) with MJD smaller than 5200 
and larger than 5200, respectively. Therefore, we said there are not trend variability in the light curve with MJD larger than 5200.

	Now, lets check whether the standard theoretical TDE model can be applied to describe the variability of SDSS J1605. 
The more recent detailed descriptions on the theoretical TDE model can be found in \citet{gr13, gm14, gn18, mg19}. The 
corresponding public codes on the theoretical TDE model can be found in TDEFIT (\url{https://tde.space/tdefit/}) and MOSFIT 
(\url{http://mosfit.readthedocs.io/}). Here, based on the more recent discussions in \citet{mg19}, and similar as what we have 
recently done in \citet{zh22, zh22b, zh22c, zh23} to model/simulate TDE expected time dependent variability, simple descriptions 
on applications of standard theoretical TDE model are given as follows by four steps.

	First, based on TDEFIT/MOSFIT provided fallback material rates $\dot{M}_{fbt}=dm/de~\times~de/dt$ for standard cases with 
a $1{\rm M_\odot}$ main sequence star tidally disrupted by a central SMBH with mass $M_{BH}=10^6{\rm M_\odot}$, standard templates 
of viscous-delayed accretion rates $\dot{M}_{at}$ (described in \citealt{mg19}) can be created by 
\begin{equation}
\dot{M}_{at}~=~\frac{exp(-t/T_{vis})}{T_{vis}}\int_{0}^{t}exp(t'/T_{vis})\dot{M}_{fbt}dt'
\end{equation}.	
And a grid of 31 $\log(T_{vis, t}/{\rm years})$ range from -3 to 0 are applied to create templates of $\dot{M}_{at}$ for each 
impact parameter $\beta$ applied in TDEFIT/MOSFIT, leading to the created templates of $\dot{M}_{at}$ including 736 (640) 
time-dependent viscous-delayed accretion rates for 31 different $T_{vis}$ of each 23 (20) impact parameters for the main-sequence 
star with polytropic index $\gamma$ of 4/3 (5/3).

	Second, simple linear interpretations are applied to determine viscous-delayed accretion rate $\dot{M}_{a}$ with input 
parameters of $\beta$ and $T_{vis}$ different from the list values in $\beta_t$ for the standard cases listed in TDEFIT/MOSFIT 
and in $T_{vis, t}$ discussed above, 
\begin{equation}
\begin{split}
\dot{M}_{a}(T_{vis}, \beta_{1}) &= \dot{M}_{at}(T_{vis1}, \beta_1) + \\
	&\frac{T_{vis}-T_{vis1}}{T_{vis2}-T_{vis1}}(\dot{M}_{at}(T_{vis2}, \beta_1)
	- \dot{M}_{at}(T_{vis1}, \beta_1))\\
\dot{M}_{a}(T_{vis}, \beta_2) &= \dot{M}_{at}(T_{vis1}, \beta_2) + \\
	&\frac{T_{vis}-T_{vis1}}{T_{vis2}-T_{vis1}}(\dot{M}_{at}(T_{vis2}, \beta_2)
	- \dot{M}_{at}(T_{vis1}, \beta_2)) \\
\dot{M}_{a}(T_{vis}, \beta) &= \dot{M}_{a}(T_{vis}, \beta_1) + \\
	&\frac{\beta-\beta_1}{\beta_2-\beta_1}(\dot{M}_{a}(T_{vis}, \beta_2)
	- \dot{M}_{a}(T_{vis}, \beta_1))
\end{split}
\end{equation}
with $\beta_1$, $\beta_2$ in the $\beta_{t}$ as the two values nearer to the input $\beta$, and $T_{vis1}$, $T_{vis2}$ in the 
$T_{vis,t}$ as the two values nearer to the input $T_{vis}$.

	Third, for a TDE with input model parameters of $M_{\rm BH}$ and $M_{*}$ different from $10^6{\rm M_\odot}$ and 
$1{\rm M_\odot}$, the actual viscous-delayed accretion rates $\dot{M}$ and the corresponding time information in observer frame 
are created from the viscous-delayed accretion rates $\dot{M}_{a}(T_{vis}, \beta)$ by the following scaling relations,
\begin{equation}
\begin{split}
&\dot{M} = M_{\rm BH,6}^{-0.5}\times M_{\star}^2\times
	R_{\star}^{-1.5}\times\dot{M}_{a}(T_{vis}, \beta) \\
&t = (1+z)\times M_{\rm BH}^{0.5}\times M_{\star}^{-1}\times
	R_{\star}^{1.5} \times t_{a}(T_{vis}, \beta)
\end{split}
\end{equation},
where $M_{\rm BH,6}$, $M_{\star}$, $R_{\star}$ and $z$ represent central BH mass in units of ${\rm 10^6M_\odot}$, stellar mass 
in units of ${\rm M_\odot}$, stellar radius in units of ${\rm R_{\odot}}$ and redshift of host galaxy of a TDE, respectively. 
And the mass-radius relation in \citet{tp96} has been accepted in the Letter for main-sequence stars.

	Fourth, the time-dependent output spectrum from the accreting process with the TDE model expected accretion rate 
$\dot{M(t)}$ can be determined by the simple black-body photosphere model as discussed in \citet{mg19},  
\begin{equation}
\begin{split}
F_\lambda(t)&=\frac{2\pi Gc^2}{\lambda^5}\frac{1}{exp(hc/(k\lambda T_p(t)))-1}(\frac{R_p(t)}{D})^2 \\
R_p(t) &= R_0\times a_p(\frac{\epsilon\dot{M(t)}c^2}{1.3\times10^{38}M_{\rm BH}/{\rm M_\odot}})^{l_p} \\
T_p(t)&=(\frac{\epsilon\dot{M(t)}c^2}{4\pi\sigma_{SB}R_p^2})^{1/4} \ \ \ \ \  
	a_p = (G M_{\rm BH}\times (\frac{t_p}{\pi})^2)^{1/3}
\end{split}
\end{equation}
where $D$ means the distance to the earth calculated by redshift $z$, $k$ is the Boltzmann constant, $T_p$ and $R_p$ represent 
the time-dependent effective temperature and radius of the photosphere, $\epsilon$ is the energy transfer efficiency smaller 
than 0.4, $\sigma_{SB}$ is the Stefan-Boltzmann constant, $t_p$ is the time information of the peak accretion rate. Then, 
time-dependent apparent magnitudes $mag(t)$ can be well determined through transmission curve of the accepted filters convolved 
with the $F_\lambda(t)$. In this letter, the commonly applied Johnson V filter is accepted.


	Therefore, based on the four steps above, apparent magnitudes $mag_m(t)$ can be created by theoretical TDE model with 
eight parameters (redshift $z=0.44096$ accepted to SDSS J1605) of central BH mass $M_{\rm BH}$, mass $M_{\star}$ and polytropic 
index $\gamma$ (4/3 or 5/3) of the disrupted main-sequence star, the impact parameter $\beta$, the viscous-delay time 
$T_{vis}$, the energy transfer efficiency $\epsilon$, and the two parameters of $R_0$ and $l_p$ to describe the photosphere. 
Moreover, there is an additional parameter of $mag_0$ to represent the contributions from stellar lights in host galaxies and/or 
from not apparent intrinsic AGN variability. Then, for the observed long-term variability of SDSS J1605, the well-known maximum 
likelihood method combined with Markov Chain Monte Carlo technique \citep{fh13} can be applied to determine the best fitting 
results. Certainly, when the procedure above applied, there is only one limitation to the model parameters, leading to the 
determined tidal disruption radius $R_{\rm TDE}~\propto~M_\star^{-1/3}M_{BH}^{-2/3}$ larger than the event horizon 
$R_{\rm s}=2GM_{\rm BH}/c^2$ of central BH.

	Before proceeding further, there is one point we should note. The circularizations in TDEs as more recently discussed in 
\citet{zo20, lo21} are not considered in this Letter. Due to probably different fallback timescales for the 
circularizations and the accretion processes, two bright phases could be expected in light curves. However, as the shown 
CSS light curve of SDSS J1605, there are no clues on two bright phases. Therefore, we mainly consider the simple case that the 
fallback timescales of the circularizations are significantly smaller than the viscous timescales of the accretion processes, 
leading the fallback materials to circularize into a disk as soon as possible.

	For SDSS J1605, the TDE model determined best-fitting results and the corresponding confidence bands by uncertainties of 
the TDE model parameters are shown in the left panel of Fig.~\ref{lmc} to the CSS V-band light curve. And right panels of Fig.~\ref{lmc} 
show the MCMC determined two-dimensional posterior distributions of the model parameters. The determined model parameters with 
$\gamma=4/3$ are as follows: $\log(M_{\rm BH,6})\sim2.16\pm0.07$, $\log(M_\star/{\rm M_\odot})\sim0.45\pm0.03$, 
$\log(\beta)\sim0.20\pm0.04$, $\log(T_{vis})\sim-0.91\pm0.14$, $\log(\epsilon)\sim-0.81\pm0.14$, $\log(R_0)\sim0.52\pm0.19$, 
$\log(l_p)\sim0.27\pm0.09$ and $\log(mag_0)=1.2668\pm0.0004$. The TDE-model expected best descriptions to the light curve provide 
clues to support a potential TDE in the BLAGN SDSS J1605. 

	The TDE model determined BH mass is about $144_{-21}+{+26}\times10^6{\rm M_\odot}$ in \obj, certainly smaller than the 
corresponding Hills limit value ${\rm 420\times10^6M_\odot}$ through the Equation (6) in \citet{yao23} after considering the larger 
stellar mass of the disrupted star in \obj, indicating the determined BH mass can be reasonably accepted in \obj. Meanwhile, as 
discussed in \citet{yao23} that TDE rate drops significantly with BH masses higher than ${\rm 10^{7.5}M_\odot}$, however massive 
stars being disrupted can lead to detectable TDEs with larger central BH masses.

\begin{table}
\caption{Parameters for the emission line components}
\begin{tabular}{cccccc}
\hline\hline
\multicolumn{6}{c}{model parameters of accretion disk model for broad H$\beta$} \\
\multicolumn{6}{c}{$r_0=63\pm14$,~ $r_1=530\pm30$,~$\sin(i)=0.27\pm0.01$,~$q=0.67\pm0.32$} \\
\multicolumn{6}{c}{$e=0.22\pm0.02$,~$\sigma_L=980\pm160{\rm km/s}$,~$\phi_0=15\pm2\degr$} \\
\hline\hline
\multicolumn{6}{c}{Gaussian parameters of narrow emission lines around H$\beta$} \\
  $\lambda_0$(\AA) & $\sigma$(\AA) & flux & $\lambda_0$(\AA) & $\sigma$(\AA) & flux \\ 
4861.5$\pm$0.3 & 2.8$\pm$0.3 & 28$\pm$3 & 4864.4$\pm$0.4 & 11.4$\pm$0.3 & 68$\pm$13 \\
5007.2$\pm$0.1 & 2.8$\pm$0.1 & 434$\pm$6 & 5011.5$\pm$0.4 & 12.5$\pm$0.4 & 245$\pm$8 \\
4845.5$\pm$2.1  & 30.9$\pm$2.4  & 479$\pm$45  & 
	4904.3$\pm$2.3  & 35.6$\pm$3.1  & 670$\pm$91  \\
\hline\hline
\end{tabular}\\
{\bf Note:}
The first (last) three columns listed in the last two rows show the central wavelength in units of \AA, the second moment in 
units of \AA~ and flux in units of ${\rm 10^{-17}erg/s/cm^2}$ for the core (extended) components of narrow H$\beta$ and 
[O~{\sc iii}]$\lambda5007$\AA, and for the two broad Gaussian components in the broad H$\beta$.
\end{table}

\begin{figure}
\centering\includegraphics[width = 8cm,height=4.5cm]{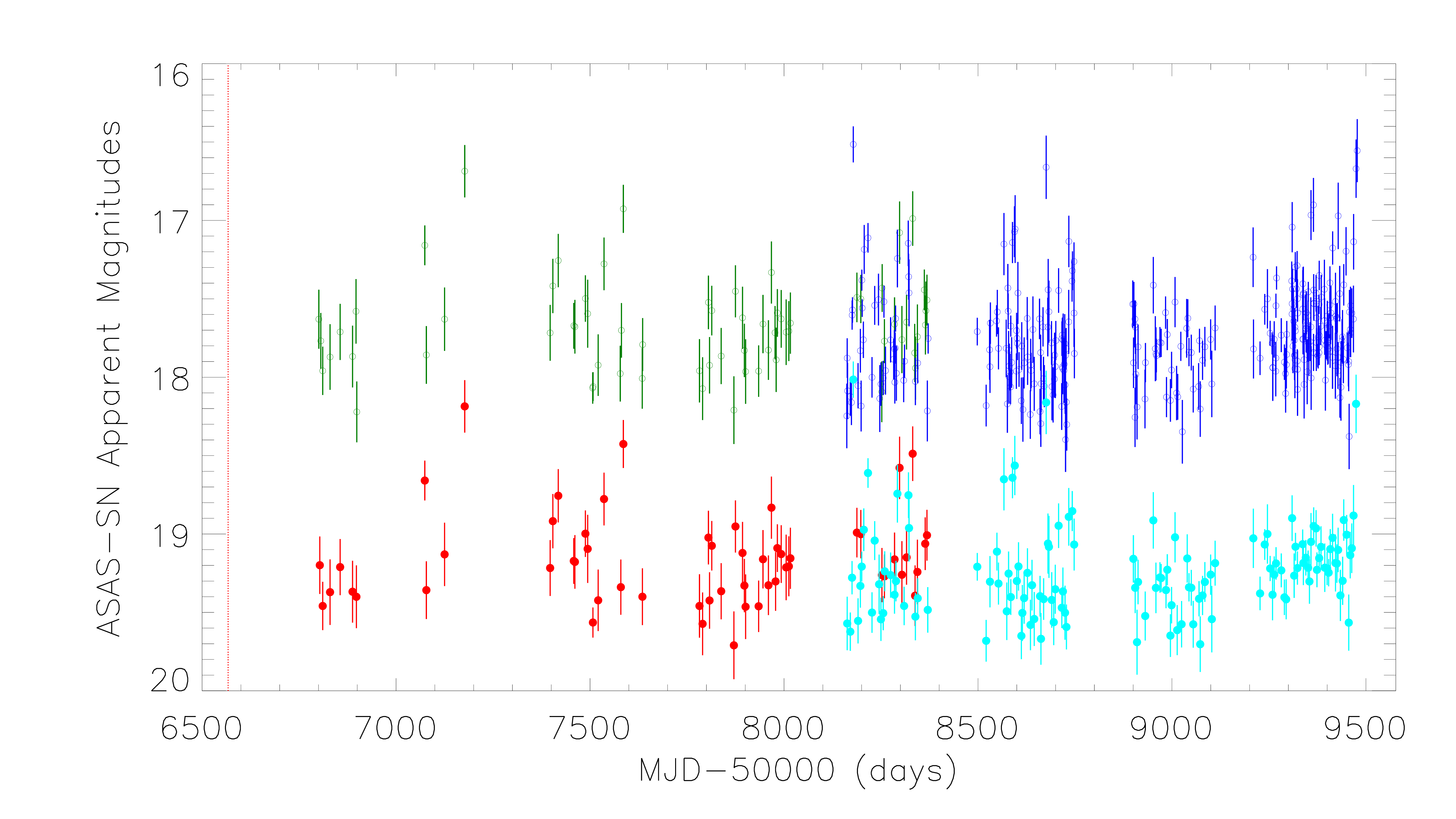}
\caption{ASAS-SN light curves of SDSS J1605. Open circles plus error bars in dark green and in blue show the V-band and g-band 
data points. The vertical dotted red line marks the position for the ending time of the CSS light curve shown in Fig.~\ref{lmc}. Solid circles plus error bars in red and in cyan show the corresponding 5days binned light curves (plus 1.5mag).}
\label{as}
\end{figure}

	Moreover, besides the CSS light curve, the 7years-long V/g-band light curves can be collected from the ASAS-SN with MJD 
from about 6800 to 9500 for SDSS J1605, and also the 5days binned light curves shown in Fig.~\ref{as}. There are not apparent trend 
variability in the ASAS-SN light curves, due to the linear function determined trend variability amplitudes about 0.07mag 
and 0.09mag in the V/g-band light curves, indicating that the bright flare around MJD$\sim3000-5000$ unique enough is not related 
to intrinsic AGN activity. Furthermore, similar as done in \citet{ht23b}, the damped random walk process \citep{kbs09, 
koz10} is applied to describe the 5days binned ASAS-SN g-band light curve with determined variability timescale about 4.5days 
in rest frame through the public JAVELIN code \citep{zk13}, leading to the expected BH mass smaller than $10^{6.2}M_\odot$ through 
the results shown in Figure~ 8 in \citet{ht23b}, which is very smaller than the TDE model determined BH mass and also very smaller 
than the virial BH mass discussed in section 3, indicating few effects of intrinsic AGN variability.

\section{Properties of emission lines in SDSS J1605}

\begin{figure*}
\centering\includegraphics[width = 18cm,height=6cm]{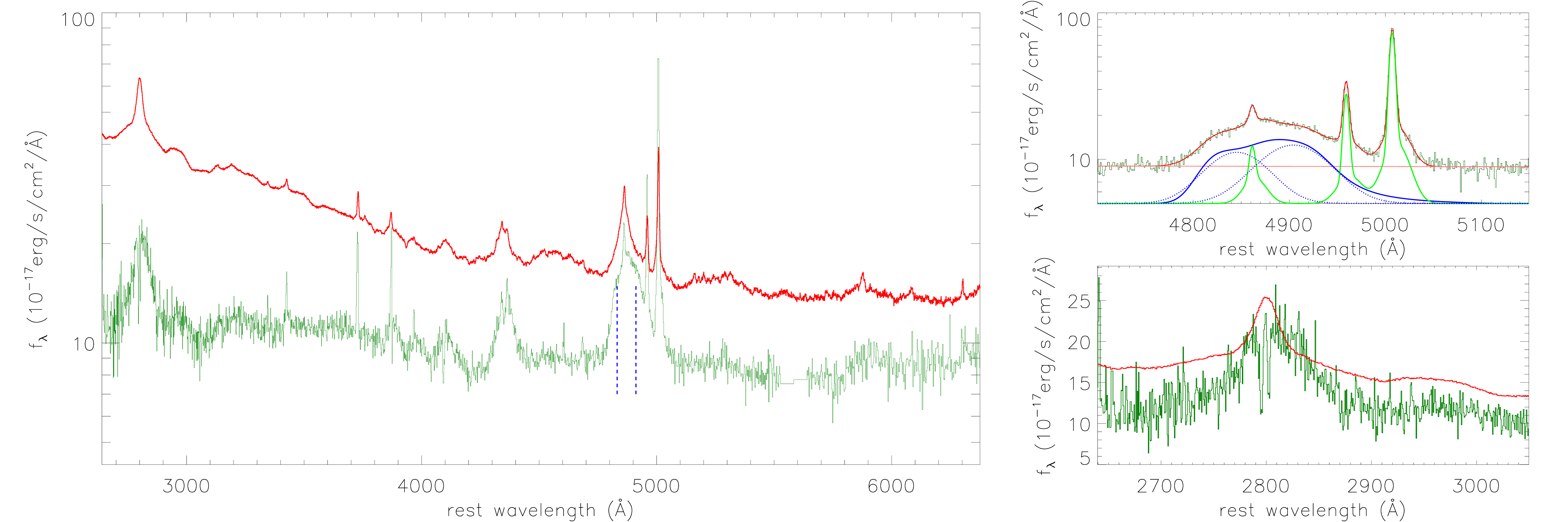}
\caption{Left panel shows the SDSS spectrum (solid dark green line) of SDSS J1605 with plate-mjd-fiberid=2524-54568-0105. Solid red line 
shows the composite spectrum of SDSS quasars, vertical dashed blue lines mark positions of two peaks (or shoulders) of the broad 
H$\beta$. Top right panel shows the best descriptions (solid red line) to the emission lines around H$\beta$ (solid dark green line). 
Solid pink line shows the determined power law component, solid blue line shows the determined broad H$\beta$ (plus 5) by the accretion 
disk model, solid green lines show the determined both core (plus 5) and extended components (plus 5) of the narrow H$\beta$ and the 
[O~{\sc iii}] doublet, dotted blue lines show the determined broad Gaussian components (plus 5) in the broad H$\beta$ 
described by broad Gaussian functions. Bottom right panel shows the emission line properties around Mg~{\sc ii}. Solid dark green line 
shows the SDSS spectrum, solid red line shows the composite spectrum of SDSS quasars.}
\label{spec}
\end{figure*}

	SDSS spectrum of SDSS J1605 (plate-mjd-fiberid=2524-54568-0105) is shown in left panel of Fig.~\ref{spec} with apparently 
broad H$\beta$ emission line with marked clear double-peaked features and also apparent broad UV Mg~{\sc ii} emission line. 
Meanwhile, the composite spectrum of SDSS quasars \citep{vb01} is also shown, to confirm the quasar-like spectroscopic 
features of SDSS J1605.

	Similar as what have done in \citet{zh21b}, multiple Gaussian functions are applied to describe the narrow emission 
lines around H$\beta$ with rest wavelength from 4700 to 5150\AA, one extended and one narrow Gaussian components for the extended 
and core components of narrow H$\beta$ and [O~{\sc iii}] doublet. And the elliptical accretion disk model discussed in \citet{el95} 
is applied to describe the double-peaked broad H$\beta$, similar as what we have recently done in \citet{zh22d}. Through the 
Levenberg-Marquardt least-squares minimization technique (the known MPFIT package), the best descriptions are shown in the top right  
panel of Fig.~\ref{spec}, with $\chi^2/dof\sim1.01$ ($\chi^2$ and $dof$ as summed squared residuals and degree of freedom). The 
measured line parameters of the narrow lines and the disk model parameters of the double-peaked broad H$\beta$ are listed in Table~1. 
Meanwhile, the spectroscopic features of the broad UV Mg~{\sc ii} emission line is shown in the bottom right panel of Fig.~\ref{spec}.

	Based on the emission line properties, among the reported TDEs candidates, SDSS J1605 has unique spectroscopic properties. 
First, among the reported TDEs candidates, only quite weak narrow forbidden emission lines can be detected in NUV and optical band. 
Even in the TDEs candidates of SDSS J0159 (a changing-look QSO) \citep{md15, lc15} and PS16DTM \citep{bn17} in the narrow line 
Seyfert I galaxy SDSS J015804.75-005221.8, there are apparent [O~{\sc iii}] emission lines, however, the flux ratios O3HB of 
[O~{\sc iii}] to narrow H$\beta$ are only 2.7 and 3.2. Similar lower O3HBs can be found in the other TDEs candidates, such as in 
XJ1500+0154 \citep{lg17} with O3HB about 1.6. However, SDSS J1605 has the strongest narrow emission lines of [O~{\sc iii}] doublet, 
with O3HB larger than 7, indicating strong and reliable central intrinsic AGN activity in SDSS J1605 through the applications of 
BPT diagram \citep{kn19, kh03}. Second, not only broad Balmer emission lines but also broad prominent He emission lines can be 
expected as spectroscopic properties of TDEs candidates, such as the cases in PS1-10jh \citep{gs12}, AT2018hyz \citep{sn20}, 
F01004-2237 \citep{ts17}, etc. However, SDSS J1605 have no apparent optical He~{\sc ii} emission lines. Third, there are no reports 
on UV broad Mg~{\sc ii} emission lines as spectroscopic properties of TDEs candidates. However, SDSS J1605 has apparent broad 
Mg~{\sc ii} emission lines. Fourth, at the late times of TDEs, the double-peaked broad Balmer and He emission lines tightly related 
to TDEs debris are becoming quite weak, such as the cases in SDSS J0159 \citep{md15}, ASASSN-14li \citep{ht16}, PTF09djl \citep{lz17} 
etc. However, SDSS J1605, its spectrum observed at MJD=54568 marked by vertical solid red line in Fig.~\ref{lmc}, has the broad 
Balmer emission lines being strong enough at late times of the expected TDE. Through the spectroscopic features of strong forbidden 
emission lines, strong broad Balmer emission lines at late times of TDE and strong UV broad Mg~{\sc ii} emission lines, 
SDSS J1605 could be a normal BLAGN as the host galaxy of an expected central TDE, not similar as the host galaxy of any one of the 
reported optical TDEs.

	If the broad H$\beta$ in SDSS J1605 was totally related to central TDE debris, the sizes of broad H$\beta$ emission regions 
could be similar as the geometric sizes of the TDE debris. Based on the commonly and well discussed accreting debris in \citet{gr13}, 
the TDE debris in SDSS J1605 at MJD=54568 have outer distance about $R_{out}~\propto~M_{\rm BH}^{1/3}t^{2/3}\sim22\ 
{\rm light-days}\sim2200{\rm R_s}$ and have inner distance about the $R_{in}\sim R_{\rm TDE}\sim2{\rm R_s}$. Meanwhile, based on the 
elliptical accretion disk model to describe the double-peaked broad H$\beta$, the model determined inner and outer boundaries in 
units of $R_s$ of the expected disk-like BLRs are about $63$ and $530$, the other five model parameters of the inclination angle 
$i$, the line emissivity slope $q$ for the broad line emissions, the local broadening velocity $\sigma_L$, the eccentricity $e$ and 
orientation angle $\phi_0$ for the disk-like BLRs are listed in Table~1. The accretion disk model determined inner and outer boundaries 
for the disk-like structure are not beyond TDE model determined $R_{out}$, indicating double-peaked features of broad H$\beta$ cannot 
provide clues to be against the potential central TDE in SDSS J1605.

	Meanwhile, through the improved R-L empirical relation for sizes of AGN BLRs in \citet{bd13}, the continuum luminosity 
($\sim3.23\times10^{44}{\rm erg/s}$) expected BLRs size is about 67light-days in SDSS J1605, three times larger than the outer 
distance of TDE debris. Then, combining with broad H$\beta$ with line width (second moment) about $2660\pm240{\rm km/s}$, the expected 
virial BH mass is about $M_{\rm BH}\sim8.83\times10^8{\rm M_\odot}$ through the virialization formula discussed in \citet{pf04, rh11}, 
similar as the virial BH mass about $9.3-10.9\times10^8{\rm M_\odot}$ in \citet{sh11} for SDSS J1605. Apparently, the virial BH mass 
is about 7times larger than the TDE model determined BH mass. Accepted TDE contributions leading to actual BLR sizes smaller than R-L 
relation estimated value, TDE model determined BH mass smaller than virial BH mass in SDSS J1605 can be reasonably explained. More 
accurate BH mass estimated by independent method in SDSS J1605 will provide clear clues to support or to be against the expected 
central TDE in SDSS J1605.

	Furthermore, besides disk-like structure discussed above, double BLRs related to a binary black hole system (BBH system) 
\citep{bb80, gd15, zh23b, zh23c} could also be applied to describe the double-peaked broad H$\beta$ in \obj. If there was a BBH 
system in \obj, through the broad H$\beta$ described by two broad Gaussian components shown as dotted blue lines in the top right 
panel of Fig.~\ref{spec} with the corresponding parameters listed in Table~1, the central BH mass for the -1030$\pm$130km/s 
blue-shifted broad component should be about $2.5_{-0.4}^{+0.5}$ times of the central BH mass for the 2570$\pm$142km/s red-shifted 
broad component. However, simply accepted the Virialization assumptions, mass ratio of the central two BHs related to the blue-shifted 
and the red-shifted broad components in the broad H$\alpha$ can be estimated as 
$\frac{f_b^{0.5}\sigma_b^2}{f_r^{0.5}\sigma_r^2}\sim0.64_{-0.23}^{+0.36}$ very smaller than the expected 2.5, with $f_b$, $f_r$, 
$\sigma_b$ and $\sigma_r$ as the emission line flux and the second moment of the blue-shifted and the red-shifted broad components 
in the broad H$\beta$. Thus the BBH system should be not favoured in \obj.

	Before ending the section, one point is noted. Besides TDE model, several another scenarios can be applied to describe 
such optical flares in AGN (or called ambiguous nuclear transient events), such as the more recent discussions in \citet{gra17, 
hn22, ht23b}. Unfortunately, due to loss of multi-band variability properties and loss of time-dependent variability properties of 
spectroscopic features in \obj, it is hard to give further discussions on the other scenarios applied in \obj. In the near future, 
based on a small sample of such flares as the one in \obj with double-peaked broad emission lines, further statistical clues 
could be provided.

	In current stage, long-term variability properties of a sample of BLAGN with double-peaked (or more complicated) broad 
emission lines are being carefully checked, after collected long-term light curves from both CSS and ZTF. We wish more TDE 
candidates can be discovered in BLAGN with double-peaked broad emission lines, and reported in the near future as soon as possible, 
after considering the basic point that BLAGN with double-peaked broad emission lines probably harboring potential TDEs candidates.

\section{Conclusions}

	A potential TDE candidate is discovered in the double-peaked BLAGN SDSS J1605, especially through TDE expected variability 
pattern in the CSS light curve. Meanwhile, none apparent trend variability in the ASAS-SN light curves afterwards provide clues to 
support the CSS flare in SDSS J1605 is unique and different enough from variability of normal AGN. Furthermore, spectroscopic results 
can be applied to determine consistency between TDE determined geometric sizes of debris and sizes of emission regions of the 
double-peaked broad H$\beta$ described by the accretion disk model, to support a central TDE in SDSS J1605. Moreover, the virial BH mass 
$\sim$7 times higher than the TDE model determined BH mass can be accepted as one another independent evidence to support a central 
TDE in SDSS J1605. Although it is still a long way from totally confirmed conclusion on the double-peaked broad lines related to TDEs, 
the results in this Letter can provide clues to detect potential TDEs in double-peaked BLAGN.

\section*{Acknowledgements}
Zhang gratefully acknowledge the anonymous referee for giving us constructive comments and suggestions to greatly 
improve our paper. Zhang gratefully thanks the kind financial support from GuangXi University and the kind financial support 
from NSFC-12173020 ad NSFC-12373014. This Letter has made use of the data from the SDSS projects. The SDSS-III web site is 
\url{http://www.sdss3.org/}. SDSS-III is managed by the Astrophysical Research Consortium for the Participating Institutions 
of the SDSS-III Collaboration. This letter has made use of the data from CSS funded by the National Aeronautics and Space 
Administration and by the U.S.~National Science Foundation.

\section*{Data Availability}
The data underlying this article will be shared on reasonable request to the corresponding author
(\href{mailto:xgzhang@gxu.edu.cn}{xgzhang@gxu.edu.cn}).

\label{lastpage}
\end{document}